\documentclass{IEEEtran}
\pdfoutput=1
\usepackage{cite}
\usepackage{amsmath,amssymb,amsfonts}
\usepackage{graphicx}
\usepackage{textcomp,nicefrac}
\usepackage{hyperref}

\def\BibTeX{{\rm B\kern-.05em{\sc i\kern-.025em b}\kern-.08em
T\kern-.1667em\lower.7ex\hbox{E}\kern-.125emX}}
\begin{document}

\title{The Mu2e crystal and SiPM calorimeter: construction status\\}
\author{Nikolay Atanov, $^{1}$, Vladimir Baranov $^{1}$, Leo Borrel $^{2}$, Caterina Bloise $^{3}$, Julian Budagov $^{1}$, Sergio Ceravolo $^{3}$, Franco Cervelli $^{4}$, Francesco Colao $^{3,5}$, Marco Cordelli $^{3}$, Giovanni Corradi  $^{3}$, Yuri Davydov $^{1}$, Stefano Di Falco $^{4}$, Eleonora Diociaiuti $^{3 }$*, Simone Donati $^{4,6}$, Bertrand Echenard $^{2}$, Carlo Ferrari $^{4}$, Ruben Gargiulo $^{3}$, Antonio Gioiosa $^{4}$, Simona Giovannella $^{3}$, Valerio Giusti $^{4,7}$, Vladimir Glagolev $^{1}$, Francesco Grancagnolo $^{8}$, Dariush Hampai $^{3}$, Fabio Happacher $^{3}$, David Hitlin $^{2}$, Matteo Martini $^{3,9}$,
Sophie Middleton $^{2}$, Stefano Miscetti $^{3}$, Luca Morescalchi $^{4}$, Daniele Paesani $^{3,}$, Daniele Pasciuto $^{4,6}$, Elena Pedreschi $^{4}$, Frank Porter $^{2}$, Fabrizio Raffaelli $^{4}$, 
 Alessandro Saputi $^{10}$, Ivano Sarra $^{3}$, Franco Spinella  $^{4}$, 
 Alessandra Taffara  $^{4}$, Anna Maria Zanetti  $^{11}$ and Ren-Yuan Zhu $^{2}$\\
 \vspace{0.5cm}
 $^{1}$  Joint Institute for Nuclear Research,Russia; \\
 $^{2}$  California Institute of Technology, Pasadena, USA;\\
 $^{3}$  Laboratori Nazionali di Frascati dell'INFN,Italy;\\
 $^{4}$  INFN---Sezione di Pisa, Italy;\\
 $^{5}$  ENEA---Frascati, Italy \\
$^{6}$  Department of Physics, University of Pisa, Italy \\
$^{7}$  Department of Civil and Industrial Engineering, University of Pisa, Italy \\
$^{8}$  INFN---Sezione di Lecce, Italy;
$^{9}$ Department of Engineering Sciences, Guglielmo Marconi University, Italy \\
$^{10}$  INFN---Sezione di Ferrara, Italy;\\
$^{11}$  INFN---Sezione di Trieste, Italy}

\maketitle

\begin{abstract}
The Mu2e experiment at Fermilab searches for the neutrino-less conversion of a negative muon into an electron, with a distinctive signature of a mono-energetic electron with energy of 104.967 MeV.  The calorimeter is made of two disks of pure CsI crystals, each read out by two custom large area UV-extended SiPMs. It plays a fundamental role in providing excellent particle identification capabilities and an online trigger filter while improving the track reconstruction, requiring  better than 10\% energy and 500 ps timing resolutions for 100 MeV electrons. In this paper, we present the status of construction and the Quality Control (QC) performed on the produced crystals and photosensors, the development of the rad-hard electronics and the most important results of the irradiation tests. Construction of the mechanics is also reported.  Status and plans for the calorimeter assembly and its first commissioning are described. \end{abstract}

\section{The Mu2e experiment and Charged Lepton Flavor Violation Processes}
The Mu2e experiment, under construction at Fermilab, will search for  the Charged Lepton Flavor Violating (CLFV) process of a muon converting into an electron in the electric field of  Al nuclei. The CLFV processes are forbidden in the Standard Model (SM) and even assuming its minimal extension, that allows neutrino oscillations, their branching ratio is completely negligible:$BR<10^{-50}$ \cite{BR}.

Any observation of CLFV processes in the muon sector will be a clear hint of New Physics.
In case of no signal event observation,  Mu2e  will set a 90\% upper limit on the ratio between the conversion and the capture rates $R_{\mu e} < 8 \times 10^{-17}$, improving the current best limit \cite{sindrumII} by four orders of magnitude.
The experimental signature searched in Mu2e is a single electron with energy slightly below the muon rest mass, that is $E_e = 104.97 $ MeV.

\begin{figure}[h!]
    \centering
    \includegraphics[width= 0.48\textwidth]{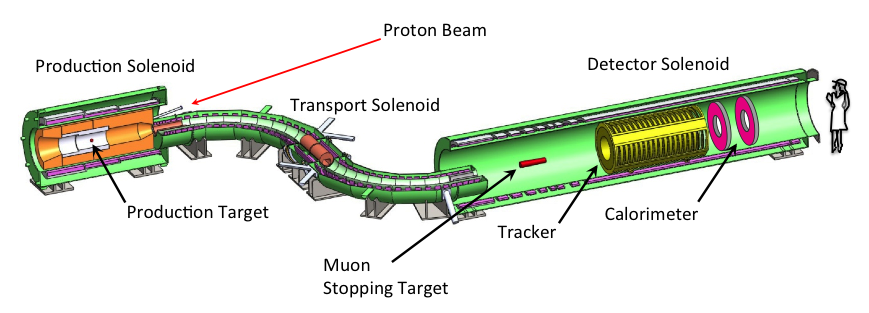}
    \caption{Layout of the Mu2e experiment: Production Solenoid, Transport and Detector solenoid are indicated in the picture. The Cosmic Ray Veto, surrounding the DS and part of the TS, is not shown.}
    \label{fig:mu2esetup}
\end{figure}

As shown in \figurename~\ref{fig:mu2esetup} the Mu2e experiment is based on a system of three superconducting solenoids to enhance the number of negative muons arriving to the Stopping Target.
A 8 GeV pulsed proton beam hitting the tungsten target inside the Production Solenoid (PS) produces mostly low momentum pions. Thanks to the graded magnetic field, particles produced forwards are reflected back towards the S-shaped Transport Solenoid (TS). This region allows having a very intense ($\sim 10^{10} \mu/s$) pure low-momentum negative muon beam at the entrance of the Detector Solenoids (DS) thanks to a middle collimator to select the particle charge. The DS houses the  stopping target (made of 37 aluminum annular 105 $\mu$m thick foils, spaced 2.2 cm apart) and the detectors - a very precise ($\sigma_p = 200$ keV  at $E_e$) straw tubes tracker \cite{tracker} and an electromagnetic calorimeter \cite{calo}. 
The tracker is composed of $\sim 20000$ low mass, very thin, straw drift tubes and will measure the charged particle momenta reconstructing their trajectories in the B-field with the detected hits.

Within its lifetime the experiment plans to collect $6 \times 10^{17}$ muon stops, necessary to reach its sensitivity goal.
To reduce Cosmic Ray contribution, the external area of the  DS, and part of the TS, are covered by a Cosmic Ray Veto (CRV) \cite{CRV} system.
 Once muons are stopped in the Al target, they create muonic atoms and then cascade to the
1S ground state, with 39\% decaying in orbit (DIO) and 61\% captured by the nucleus.
In the last case, due to the occuring nuclear processes, low energy protons, neutrons and photons are emitted, originating a large neutron flux as well as a large ionizing dose  in the detectors.

\begin{figure}[h!]
    \centering
    \includegraphics[width = 0.49 \textwidth]{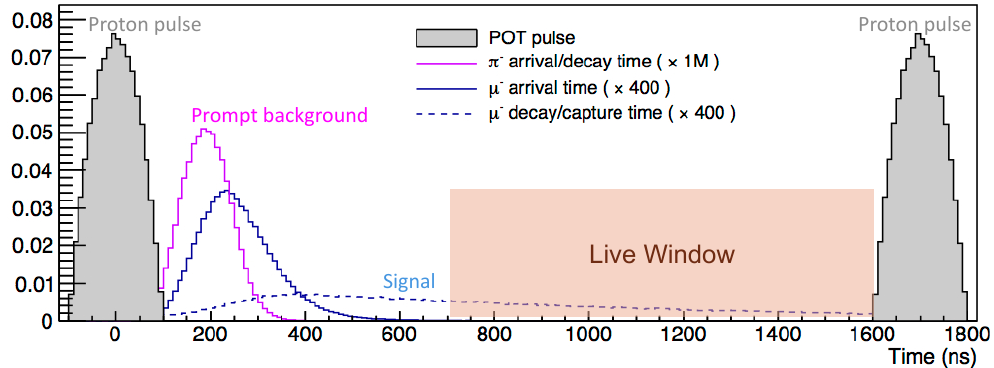}
    \caption{Timing structure of the Mu2e beam}
    \label{fig:TempoFascio}
\end{figure}

 To improve the current best limit by four orders of magnitude, Mu2e differs from earlier muon-to-electron conversion experiments in three major ways:
 \begin{itemize}
     \item the muon beam intensity is 10000 times greater than those of the previous experiments;
     \item the presence of the TS, besides providing muon sign-selection, suppresses the neutral particles contribution at the entrance of the DS, allowing an efficient muon transport to the stopping target;
     \item the pulsed structures of the beam and a delayed acquisition window: in order to suppress the prompt background, the muons hitting the stopping target are intended to be distributed in a narrow time burst ($<$ 200 ns) as shown in Figure~\ref{fig:TempoFascio}, with a bunch separation of  $\sim 1.7 \mu s$ (i.e. larger than 826 ns, the muonic aluminum lifetime). Their decay products are observed only 700 ns after the proton arrival to make the prompt background negligible. These choices were guided by the observation that the result of the SINDRUM II experiment was ultimately limited by the need of suppressing the prompts.
 \end{itemize}

The data-taking plan has been organized into two  periods (Run I and RUN II), well separated by a two-year-long shutdown for the installation of PIP-II, the linac for the DUNE experiment. In Run I only 10\% of the total number of protons on target (POT) will be produced. 75\% of these POTs will be delivered with a low  intensity proton beam with a mean intensity of $1.6 \times10^7$ protons/pulse,  while  the remaining  25\% will be delivered in the high intensity  mode ($3.9 \times 10^7$ protons/pulse).

\section{The Mu2e electromagnetic calorimeter}
To validate the charged particle  reconstructed by the tracker, the Mu2e calorimeter \cite{fTDR} provides information about its energy, timing and position, adding particle ID
capabilities to reject muons and antiprotons interactions mimicking the signal. In
addition, the calorimeter is required to be also fast enough to provide a tracker independent software trigger and help the tracks seeding \cite{seed}.

To accomplish these requirements the calorimeter has to maximise the acceptance for $\sim$ 105 MeV/c Conversion Electron (CE) tracks, operate in vacuum, survive in the ``harsh'' radiation environment and satisfy the experimental requirements discussed below.

\subsection{Experimental requirements}
To fulfill the previous stated tasks, simulation guided us in defining the reconstruction requirements for 105 MeV 
electrons, that are summarized by this short list: 
\begin{itemize}
    \item an energy resolution better than $\sigma_E/E = \mathcal{O}(10\%)$, to reach a rejection factor at the level of 200 between CE and the $\sim 40$ MeV energy deposit from 105 MeV/c cosmic ray muons mimicking the signal \cite{Run1}; 
    \item a timing resolution better than $\sim 0.5$ ns, to ensure that the energy depositions in the calorimeter are in time with the conversion electrons reconstructed by the tracker and also improve the PID;
    \item a position resolution $\sigma_{r,z} <  1$ cm, to match the position of the energy deposit with the extrapolated trajectory of a reconstructed track; 
    \item ability to survive the high radiation environment, maintaining its functionality for radiation exposures up to $\sim 15$ krad/year in the hottest regions and for a neutron flux equivalent to $10^{12}$ n$_{1MeV}$/cm$^2$ / year, inside an evacuated region ($10^{-4}$ Torr) of the DS that provides 1 T axial magnetic field; 
    \item a fast enough response in order to handle the experimental high rate ($\tau < $ 40 ns); 
    \item a temperature and gain stability within $\pm 0.5\%$, not to deteriorate the energy
resolution;
    \item reliability and redundancy to operate in vacuum for one year without any interruptions.
\end{itemize}           
%(a) an energy resolution better than $\sigma_E/E = \mathcal{O}(10\%)$ , to distinguish the CE from the $\sim 40$ MeV energy deposit from 105 MeV/c cosmic ray muons mimicking the signal \cite{Run1}; (b) a timing resolution better than $\sim 0.5$ ns, to ensure that the energy depositions in the calorimeter are in time with the conversion electrons reconstructed by the tracker and also improve the PID; (c) a position resolution $\sigma_{r,z} <  1$ cm, to match the position of the energy deposit with the extrapolated trajectory of a reconstructed track; (d) ability to survive the high radiation environment, maintaining its functionality for radiation exposures up to $\sim 15$ krad/year in the hottest regions and for a neutron flux equivalent to $10^{12}$ n$_{1MeV}$/cm$^2$ / year, inside an
%evacuated region ($10^{-4}$ Torr) of the DS that provides 1 T axial magnetic field; (e) a fast enough response in order to handle the experimental high rate ($\tau < $ 40 ns); (f) a temperature and gain stability within $\pm 0.5\%$, not to deteriorate the energy
%resolution and (g) reliability and redundancy to operate in vacuum for one year without any interruptions.

\subsection{Technical choice}
In 2015, after a long R\&D phase \cite{rd1} \cite{rd2} carried out to better define the detector, the final design of the Mu2e calorimeter was decided: a high quality undoped CsI crystal calorimeter with Silicon Photomultipliers (SiPMs) readout and with a geometry organised in two annular disks. 

Indeed, undoped CsI represents the best compromise between cost, reliability, performance and radiation hardness, providing a fast emission time ($\tau = 30$ ns) and a sufficiently high Light Yield ($\sim 2000$ $\gamma$/MeV). 
To well match the scintillation emission of 310 nm to the SiPM Photon Detection Efficiency, UV extended Hamamatsu SiPMs with a front window made of silicon resin were selected.
To operate in vacuum and minimize outgassing contributions, the crystal-SiPM
coupling was done without any optical grease.

\begin{figure}[h!]
    \centering
    \includegraphics[width = 0.49 \textwidth]{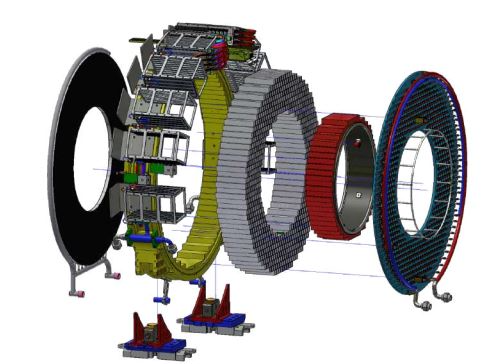}
    \caption{ Breakout of calorimeter mechanical components}
    \label{fig:caloDesign}
\end{figure}
In Figure \ref{fig:caloDesign}, the design of the calorimeter is shown: two annular disks with an inner (outer) radius of 35 cm (66 cm) and a relative distance of 70 cm, corresponding to half pitch of the helical CE trajectory. Each disk is composed of 674 square based scintillating crystals of $3.4 \times 3.4\times 20$ cm$^3$. Even if the crystal length is only 10 $X_0$, it is sufficient to contain the 105 MeV electron showers since the CEs impinge on the calorimeter surface with a $\sim 50^{\circ}$ angle.
To improve reliability, light collection and resolution, each crystal is readout by two custom design SiPMs array. Each one is a parallel configuration of two series of three $6 \times 6$ mm$^2$ monolithic sensors. The array dimension where selected to maximize the light collection, obtaining an active area of $1.8 \times 1.2$ cm$^2$ and keeping a small total capacitance - thanks to the parallel configuration-  to achieve a signal width smaller than 200 ns.\\
Each crystal is wrapped with a 150 $\mu$m thick foil of Tyvek\textregistered. The Front-End Electronics (FEE) is mounted on the rear side of each disk on the SiPMS pins, while voltage distribution, slow control and digitizer electronics are housed behind each disk in custom crates. Each FEE/SiPMs/Crystal system has its own independent powering and readout channel.

\subsection{Mechanics}
The calorimeter mechanical structure was designed to support the layout of the crystals
by piling them up in a self-standing array organized in consecutive staggered rows. Each crystal array is supported by two coaxial cylinders. The inner cylinder must be as thin and light as possible in order to minimize the passive material in the region
where spiraling background electrons are concentrated. 
 The outer cylinder is as robust as required to support the load of the crystals (700 kg). Each disk has two cover plates. The plate facing the beam is made of carbon fiber to minimize the degradation of the electron energy, while the back plate should also be robust to support the SiPMs, the FEE and the SiPM cooling lines and it is therefore made of Polyether ether ketone (PEEK). The crystal arrangement is self-supporting, with the load carried primarily by the outer ring.\\
 The heat generated by SiPMs, FEE and read out electronics must be removed within temperature values acceptable for the correct operation of each device. Furthermore, the difficult access to components requires a cooling system free of fault and maintenance needs for at least one year. The
cooling system has to maintain SiPM temperature at $ \sim -10^{\circ}$ C to minimize the dark current: this is obtained by choosing 3M NOVEC 123 HFE 1700) as refrigerating fluid, circulating at $\sim -15 ^{\circ}$ C.

\subsection{Electronics}
\begin{figure}[h!]
    \centering
    \includegraphics[width=.5\textwidth]{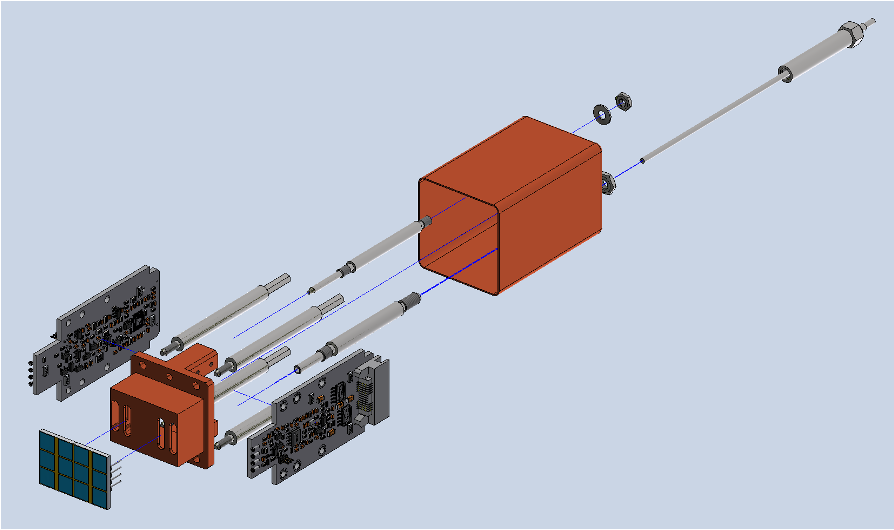}
    \caption{Exploded view of the ROU components.}
    \label{fig:ROU}
\end{figure}

The electronics is based on analog FEE cards directly connected to the SiPM pins and by a digital readout part distributed on crates surrounding the disk. As show in \figurename~\ref{fig:ROU}, the FEEs are mounted on a copper support where the SiPMs are glued. In the following the ensemble of these components will be call Read-Out Units (ROU).
FEEs provide amplification and shaping of the signals, and a local high voltage (HV) regulation for independent control of the bias voltages (through high-voltage ADCs and DACs). Readout of current and temperature sensors are also provided.

\begin{figure}[h!]
    \centering
    \includegraphics[width=0.5\textwidth]{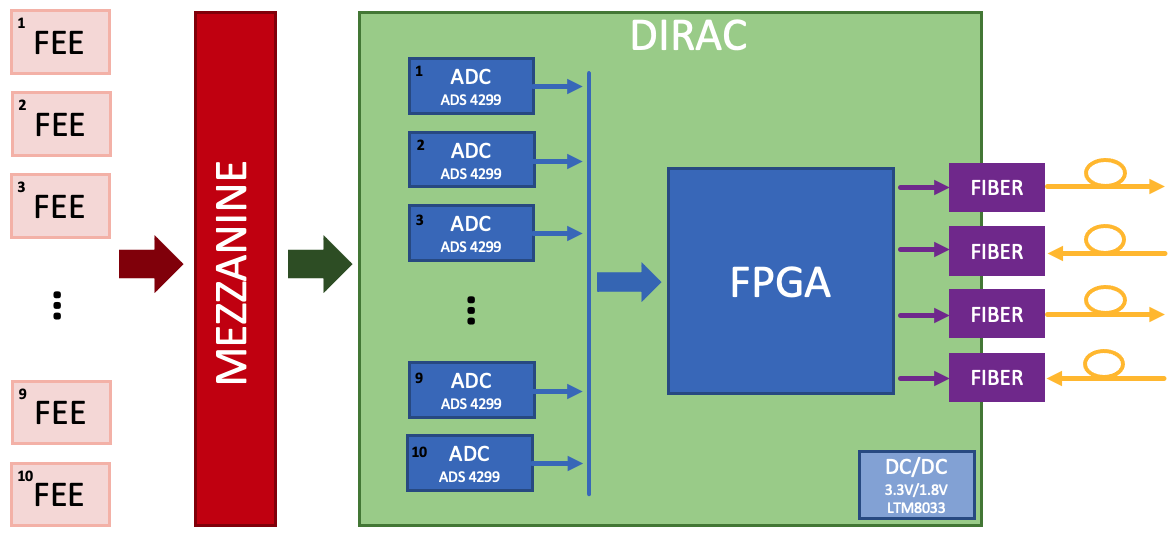}
    \caption{Block diagram for the Calorimeter Waveform Digitizer board, including also the MZB and the Amp-HV chips. The main components used are shown.}
    \label{fig:mbdir}
\end{figure}

Data Acquisition and digitization are handled by a Mezzanine Board - controlling the HV and  monitoring the current and temperature of SiPMs and a readout- and a Digitizer Readout Controller (DIRAC) board - to perform the zero-suppression and sample the signals with 5 ns binning.  Each digital board handles 20 channels. A schematic view of the FEE-MZB-DIRAC chain is shown \figurename~\ref{fig:mbdir}.

\subsection{Calibration and performance monitoring}

\begin{figure}[h!]
    \centering
    \includegraphics[width = .18 \textwidth]{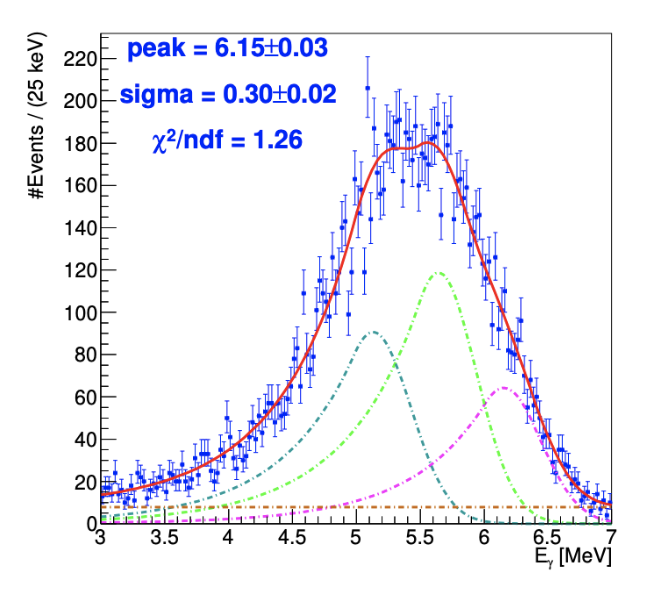}    \includegraphics[width = .3 \textwidth]{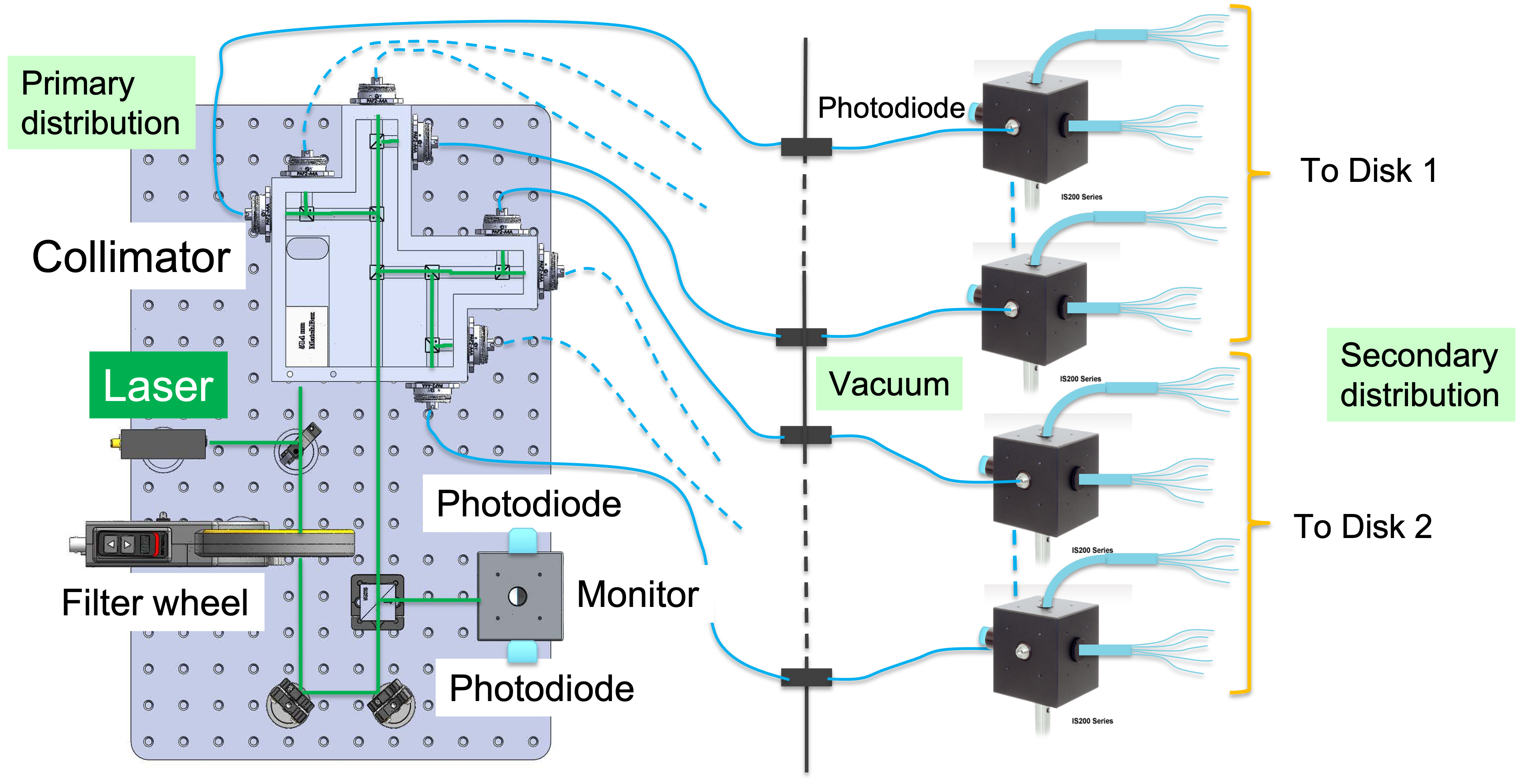}   
    \caption{Left : Energy spectrum for a crystal irradiated with 6.13 MeV photons from the  $^{16}O$. Right: primary and secondary distribution laser system.}
    \label{fig:calib}
\end{figure}

The high sensitivity required by the Mu2e experiment implies a special care in detector calibration to avoid any related systematic effects. 
A liquid radioactive (\figurename~\ref{fig:calib} Left) source will provide an absolute energy scale and a fast response equalization between crystals. This system is similar to the one developed for the BaBar calorimeter \cite{babarRadio}:
a 6.13 MeV photon line is obtained  from a short-lived $^{16}$O transition. The decay chain comes from a Fluorinert$^{TM}$  coolant liquid (FC-770) that is activated by fast neutrons produced by a DT-generator.
A continuous monitoring of  gains and time offset for each channel is instead provided by a laser monitoring system  (\figurename~\ref{fig:calib} Right): the laser light is distributed to each channel via a primary and a secondary distribution system, through an optical fiber whose final support is inserted inside the ROU structure as shown in \figurename~\ref{fig:ROU}

\section{Calorimeter performance}
\begin{figure}[h!]
    \centering
    \includegraphics[width = 0.24 \textwidth]{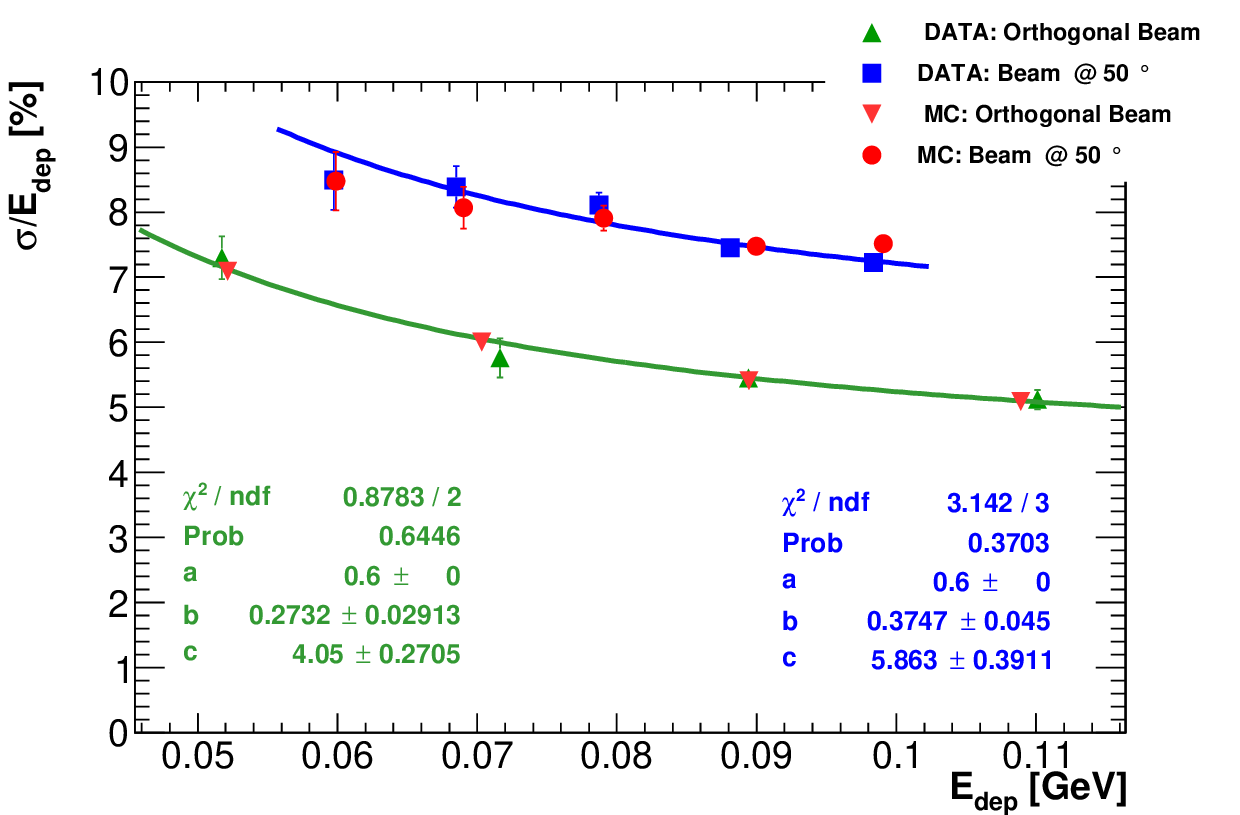}
    \includegraphics[width = 0.24 \textwidth]{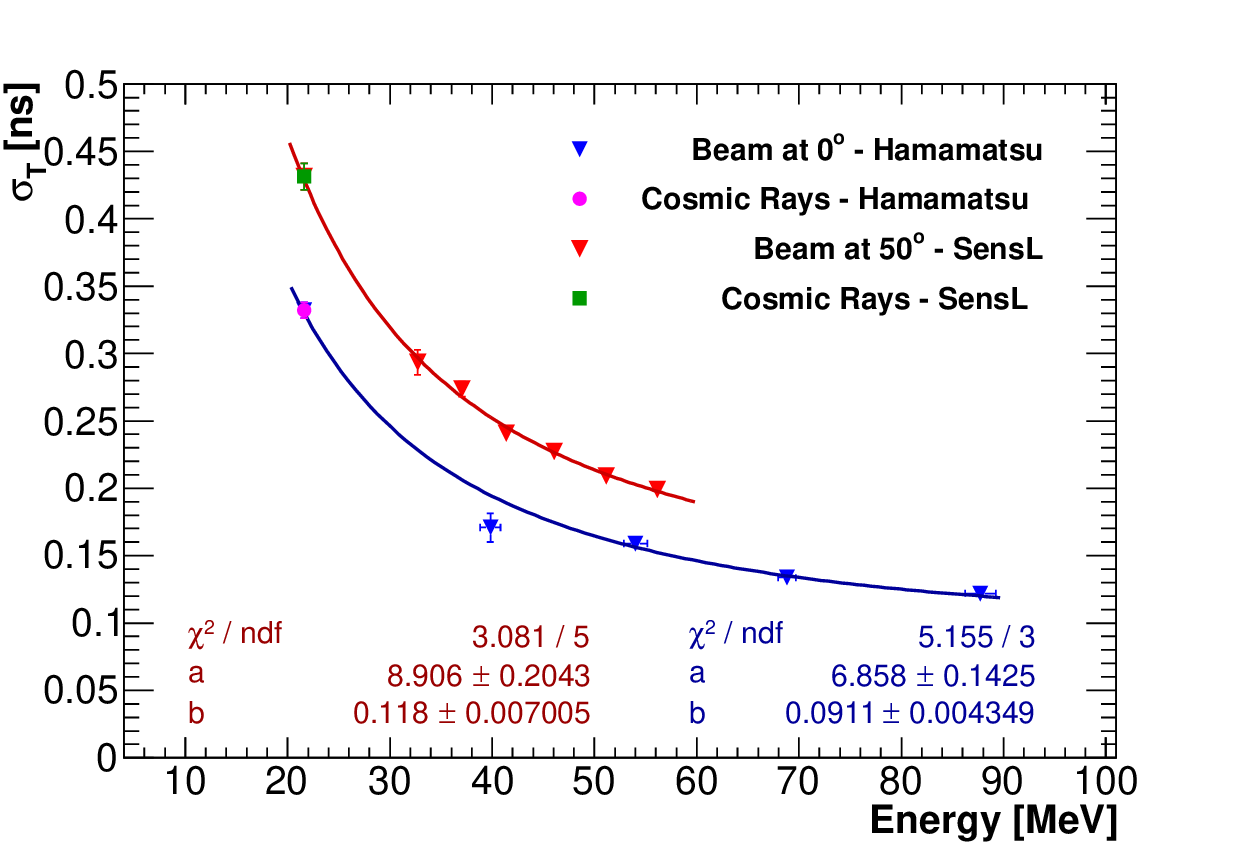}
    \caption{Energy (Left) and time (Right) resolution obtained with  the electron beam test performed at BTF. }
    \label{fig:tb}
\end{figure}

To verify the selected design before starting the massive production of the whole calorimeter components, a large size prototype was build and exposed to an electron beam at the Beam Test Facility of the National Laboratories of Frascati. The Module-0 is composed of 52 undoped CsI crystals read out by 102 SiPM connected to FEE boards. Its mechanics was designed to resemble as much as possible that of the final calorimeter, allowing the test assembly procedures and cooling. 
A detailed description of the setup and the main results (reported in \figurename~\ref{fig:tb}) obtained during the beam test can be found in \cite{tb}. 
For what concerns the energy resolution, the function used to fit the data is:
\begin{equation}
    \frac{\sigma_E}{E}= \frac{a}{\sqrt{E[GeV]}}\oplus \frac{b}{E[GeV]} \oplus c
\end{equation}
the dominant term was the constant term $c$ due to leakage, the stochastic term $a$ was consistent with a light yield of $\mathcal{O}(20$ pe/MeV/SiPM), the noise term $b$ was dominated by an electronic noise of  $\mathcal{O}(400)$ keV/channel and a coherent noise related to the used digitizers.
Runs at angle were also performed to study the performance expected with CE; due to longer clusters and higher transversal leakage the linear and constant term increased. An energy resolution better than 5 (7.5)\% 
was achieved at 100 MeV for runs at normal (at 50$^{\circ}$) incidence.

\begin{figure}[h!]
    \centering
    \includegraphics[width =0.44\textwidth]{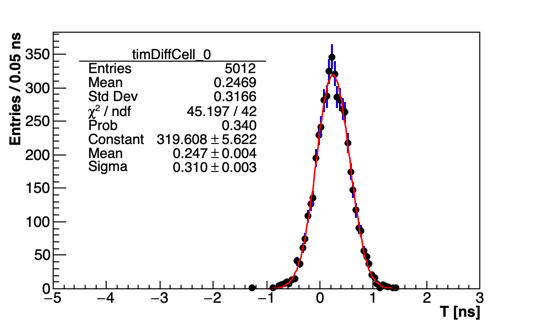}
    \caption{Timing resolution of a single crystal of the large area prototype}
    \label{fig:mod0cassone}
\end{figure}

 Time resolution was evaluated as the difference between the time of the two SiPM coupled to the same crystal, obtaining a timing resolution better than 200 ps at 100 MeV. 
  The fit function used for the time resolution is:
 \begin{equation}
      \sigma_{T} = \frac{a}{E[GeV]}\oplus b
 \end{equation}
 
 where $a$ is proportional to the emission time constant of the undoped CsI and $b$ represents the additional contribution due to the readout electronics.
In the years following the test beam, the Module-0 has continuously operated to study the behaviour in vacuum, at low temperature as well as for carrying out vertical slice tests of increasing complexity, at a Cosmic Ray test stand where the timing calibration algorithms were improved. The timing resolution obtained subtracting time of two SiPMs reading the same crystal is reported in \figurename~\ref{fig:mod0cassone}.

\section{Calorimeter production status}

\subsection{Crystal and SiPMs characterization}
\begin{figure}[h!]
    \centering
    \includegraphics[width = .48 \textwidth]{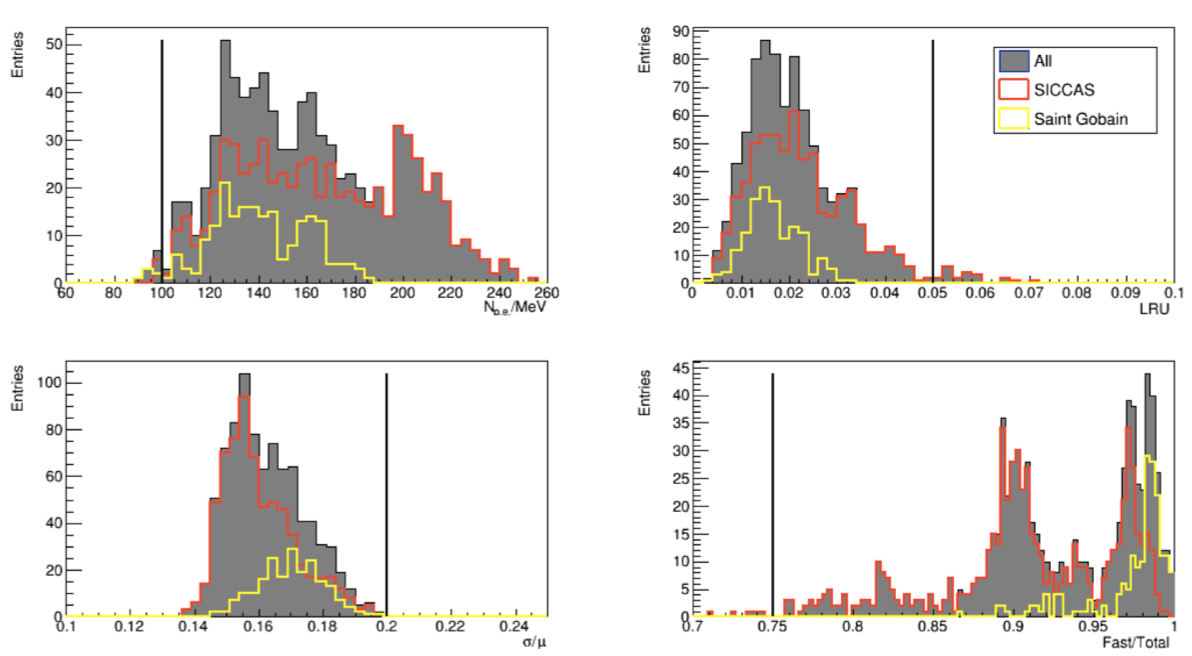}
    \caption{Number of photoelectrons (top left), LRU (top right), energy resolution (bottom left) and fast over total ration (bottom left) distributions for Saint Gobain (yellow) and Siccas (red). The comulative distributions are reported in grey. The black lines represent the experiment requirements.
}
    \label{fig:Cry_QC}
\end{figure}

\begin{figure}[h!]
    \centering
    \includegraphics[width = .48 \textwidth]{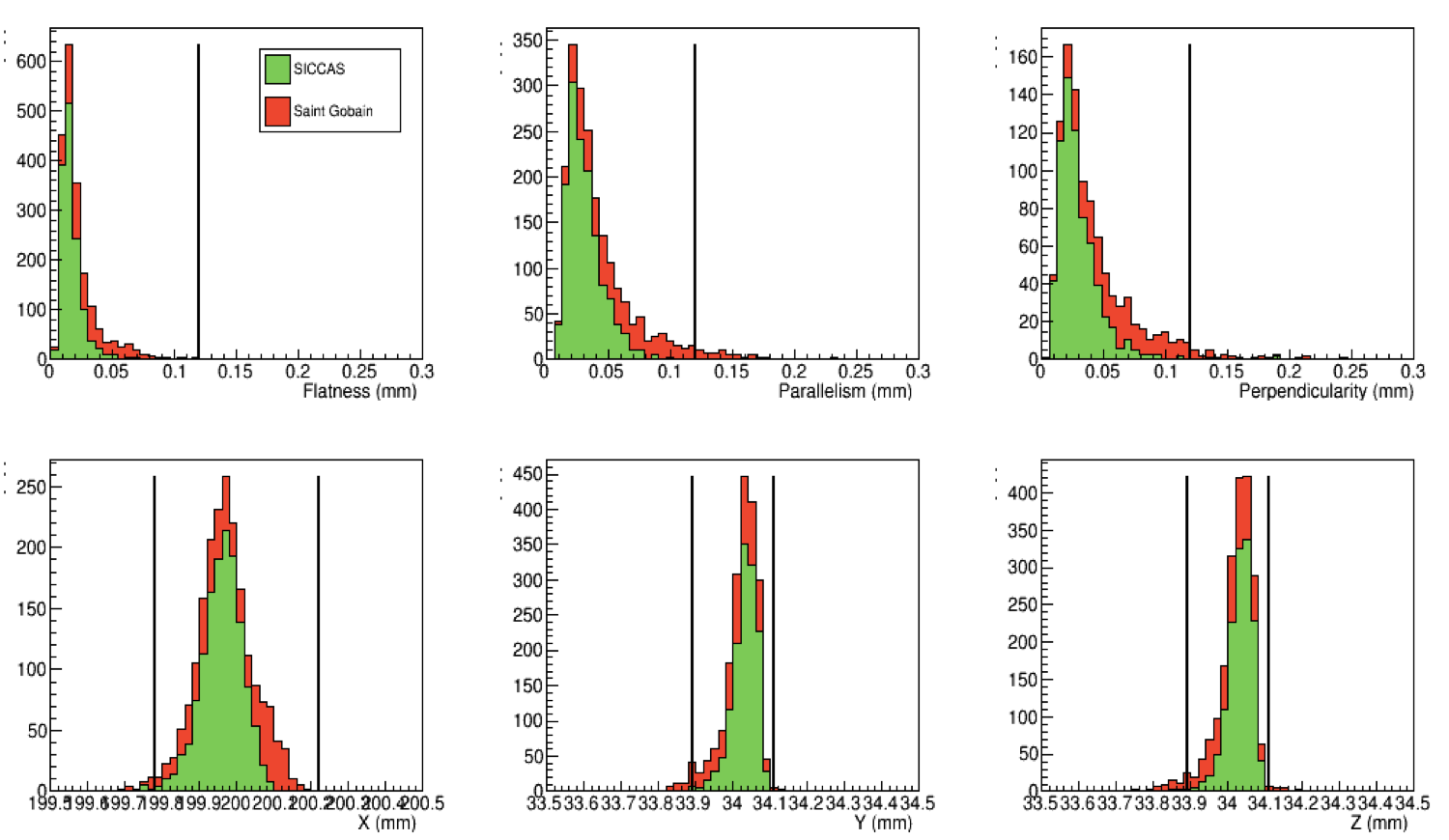}
    \caption{Results of the CMM measurements on the production crystals. Flatness (top left), parallelism (top center) and perpendicularity (top right) and the X (bottom left), Y (bottom center) and Z (bottom right) dimensions. The black lines represent the requirements.}
    \label{fig:Cry_CMM}
\end{figure}

The physics requirements previously discussed were used to define a set of technical specifications for the undoped CsI crystals and SiPMs. The selection criteria to accept crystals were:  \textit{(i)} a Light Yield LY$>$ 100 p.e./MeV; \textit{(ii)} an energy resolution better than 20\% at 511 keV; \textit{(iii)} a Longitudinal Response Uniformity (LRU), defined as the RMS of the LY measured in eight or more points along the longitudinal axis, below 5\%; \textit{(iv)} a ratio between fast and total light yield components (F/T) above 75\%.
1450 Undoped CsI crystals were produced by Siccas \cite{SICCAS} and Saint Gobain \cite{SG}. Results of the optical properties measurements are reported in \figurename~\ref{fig:Cry_QC}.
As shown in \figurename~\ref{fig:Cry_CMM} top, the results from the optical measurements were well in agreement with the specification for both vendors but Sain Gobain evidenced  some difficulties in matching the 100 $\mu$m precision on the dimensions. A few, randomly selected, crystals were exposed to neutron and photon flux, the Light Yield degradation observed remained between the limit set by the required specifications.

The specification for the Mu2e photosensors were:
\textit{(i)} a Photon Detection Efficiency (PDE) larger that 20\% at 315 nm;
\textit{(ii)} a gain larger than $10^6$ at the operational voltage $V_{op} = V_{br} + 3$ V;
\textit{(iii)} a recovery time ($\tau$) smaller than 100 ns for each of the $6\times6$ mm$^2$ cell, when measured on a load larger than 15 $\Omega$;
\textit{(iv)} a maximum acceptable operation voltage spread among the 6 SiPMs cells of $\pm$ 0.5\%;
\textit{(v)} a maximum acceptable dark current ($I_{dark}$) spread at $V_{op}$ among the 6 SiPMs cells of $\pm 15$ \%;
\textit{(vi)} $I_{dark}$ smaller than 10 mA at operation voltage and a gain reduction smaller than a factor of 4 while irradiating SiPMs up to $3\times 10^{12}$ n$_{1MeV}$/cm$^2$ at 20$^{\circ}$ C;
\textit{(vii)} a Mean Time To Failure (MTTF) better than a million hours while operating at 0$^{\circ}$ C.

\begin{figure}[h!]
    \centering
    \includegraphics[width = .45 \textwidth]{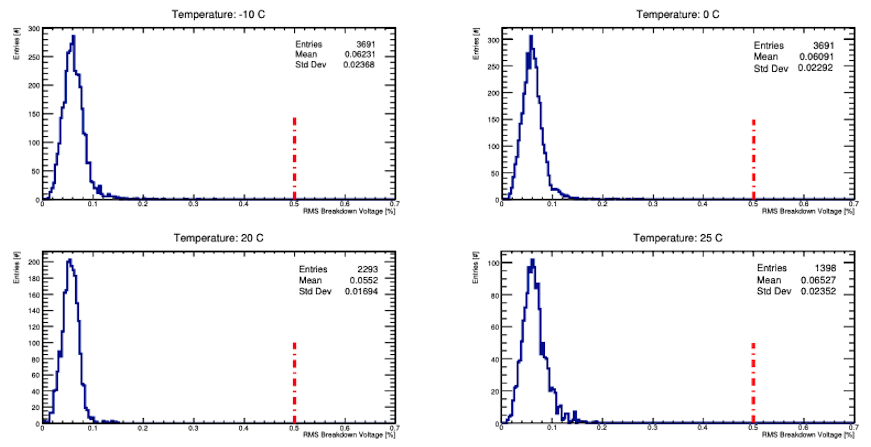}
    \caption{RMS of the Breakdown voltage test of each cell of the SiPMs. The results are reported for the three different temperature tested. The red lines represent the limits on the acceptable values.}
    \label{fig:SiPM_QC}
\end{figure}

As already stated 4000 Mu2e SiPMs (considering also spares) were produced by Hamamatsu \cite{Ham}. Each monolithic cell of the SiPMs was tested to measure the operational voltages ($V_{op}$), the dark current at $V_{op}$ and the product of Gain$\times$PDE in a custom test stand. Some of the results obtained can be seen in \figurename~\ref{fig:SiPM_QC}.

\begin{figure}[h!]
    \centering
    \includegraphics[width = .44 \textwidth]{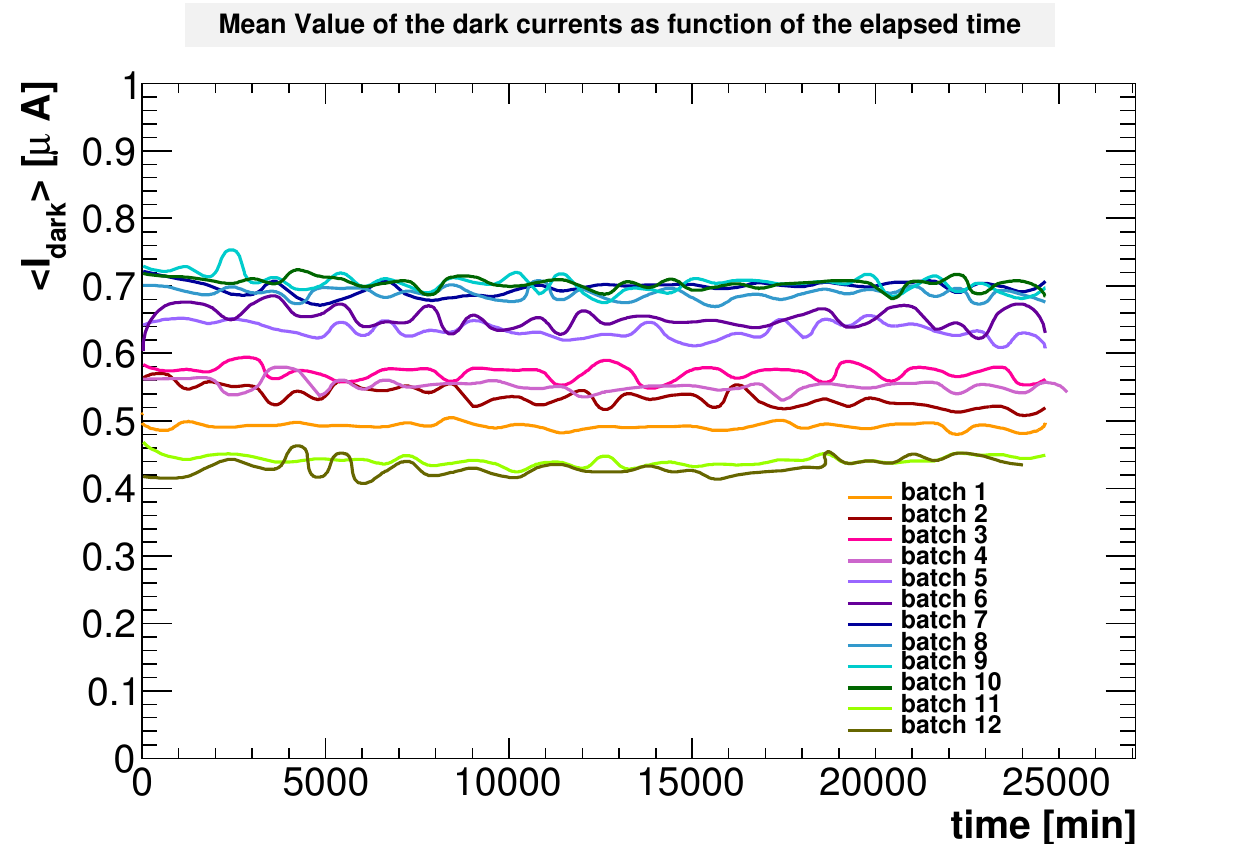}
    \caption{Mean value of the currents of 15 SiPMs as a function of the elapsed time during the MTTF test. The different line colors refers to different batches tested.}
    \label{fig:MTTF}
\end{figure}

A random sample of SiPMs was selected to be irradiated with neutron fluence up to $\sim 10^{12}$ n$_{1MeV}$/cm$^2$ while other SiPMs were selected to evaluate their Mean Time To Failure. To test an MTTF at the level of 10$^6$ hours, the SiPMs under test were kept in operation in a dedicated box at 65 $^{\circ}$ C for 342 hours. At the end of the test, no failures were observed thus allowing us to estimate a MTTF$>10^7$. Results of the test are reported in \figurename~\ref{fig:MTTF}.

\subsection{Front-End Electronics}

The FEE boards have to provide: \textit{(i)} a signal rise time of $\sim$ 25 ns to allow an appropriate time reconstruction; \textit{(ii)} a rate capability up to 1 MHz and a short fall time; \textit{(iii)} a radiation-hardness for up to 100 krad and  $10^{12}$ n$_{1MeV_{eq}}$/cm$^2$), \textit{(iv)} a programmable bias voltage up to 200 V via a 12-bit DAC, \textit{(v)} the possibility to set SiPM bias and to monitor current and temperature  via a 12-bit ADC.

\begin{figure}[h!]
    \centering
    \includegraphics[width = 0.48 \textwidth]{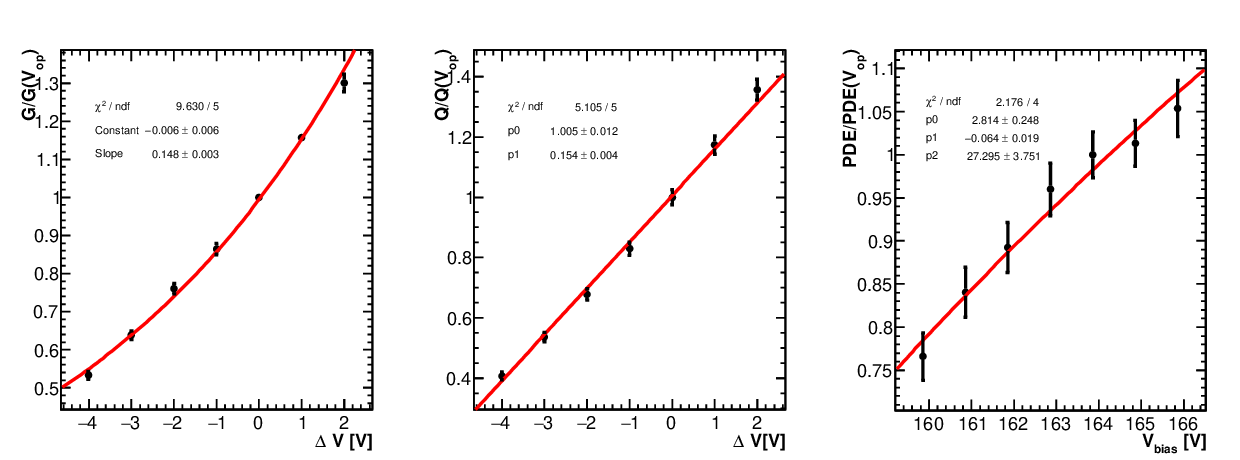}
    \caption{Example of the results of a 7 points scan for gain, collected charge and PDE. All values are normalised to the results at Vop. G and Q are show as a function of the overvoltage, while the PDE vs $V_{bias}$}
    \label{fig:ROUproperties}
\end{figure}

The FEE boards were produced by ARTEL \cite{artel} with a negligible rejection factor albeit all 2500 boards underwent a burn-in test at 65 $^{\circ}$ C in a climatic chamber in JINR (Dubna, Ru), followed by a calibration phase for both HV, gain and differential linearity parameters. 
After assembling the FEE on the copper support for the Mu2e SiPMs, a characterization of the entire ROU is performed to evaluate the Gain, charge and PDE dependence on the bias voltage (\figurename~\ref{fig:ROUproperties}) in a region around~(-4 V $\div$ +2 V)  the operational voltage.

\begin{figure}[h!]
    \centering
    \includegraphics[width = 0.48 \textwidth]{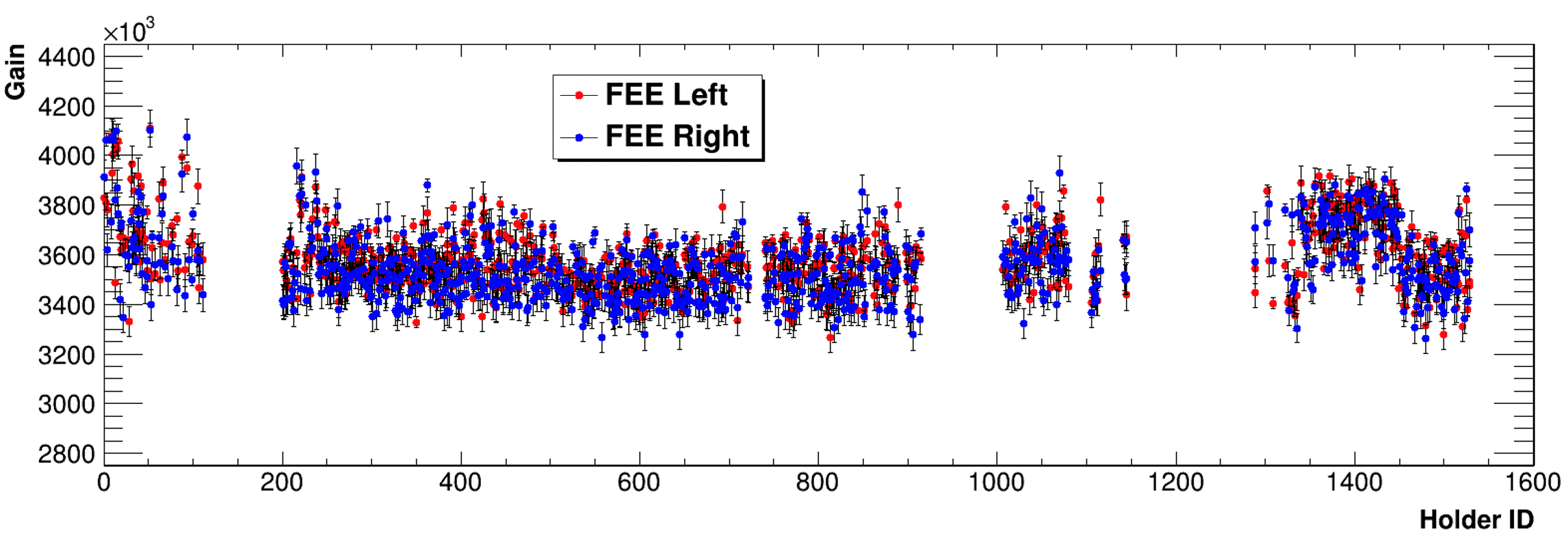}
    \caption{Gain distribution with respect to the Holder ID. The gain for the left and right SiPM-FEE couples are plotted separately for each holder.}
    \label{fig:ROUresult}
\end{figure}

At the moment of writing 1150 out of 1500 ROUs (1348 for the disks, plus spares) have been already assembled and more than 1000 fully characterized. In \figurename~\ref{fig:ROUresult} the gain at operational voltage for all the tested ROUs is reported.
\subsection{Dark current control at the end of data taking} \label{sub:idark}

A small sample of 50 SiPMs - randomly selected during the production - were exposed unbiased to a neutron fluence of  $\sim 10^{12}$ n$_{1MeV}$/cm$^2$  at the EPOS~\cite{EPOS} facility of  Helmholtz-Zentrum Dresden-Rossendorf (HZDR). The $I_{dark}$ increase was measured once the SiPMs were brought back to Fermilab. 

To better characterize the response of irradiated SiPMs, and further verify the final calorimeter operational temperature, a total of 35 SiPMs were irradiated at the Enea Frascati Neutron Generator (FNG) \cite{FNG}, using a 14 MeV neutron gun based on T(d,n)$\alpha$ fusion with neutron fluences ranging from $5 \times 10^{10}$ up to $1 \times 10^{12}$.\\
After the irradiation, the measurement of I-V  curves  was performed at different temperatures ($+15 ^{\circ}$C, $+10 ^{\circ}$C, $+5 ^{\circ}$C, $0 ^{\circ}$C and $-5 ^{\circ}$C) to evaluate the SiPM operational voltage and the dark current at $V_{op}$.
\begin{figure} [h!]
    \centering
    \includegraphics[width = .24 \textwidth]{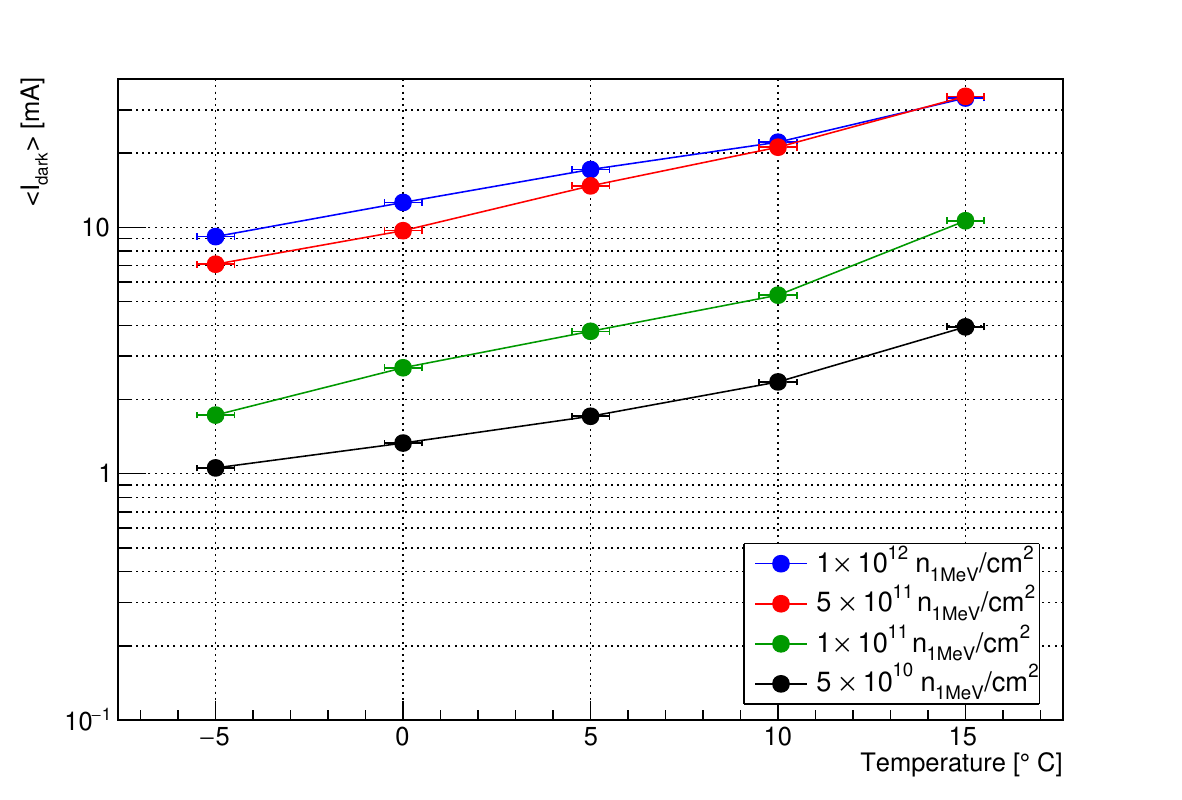}  \includegraphics[width = .24 \textwidth]{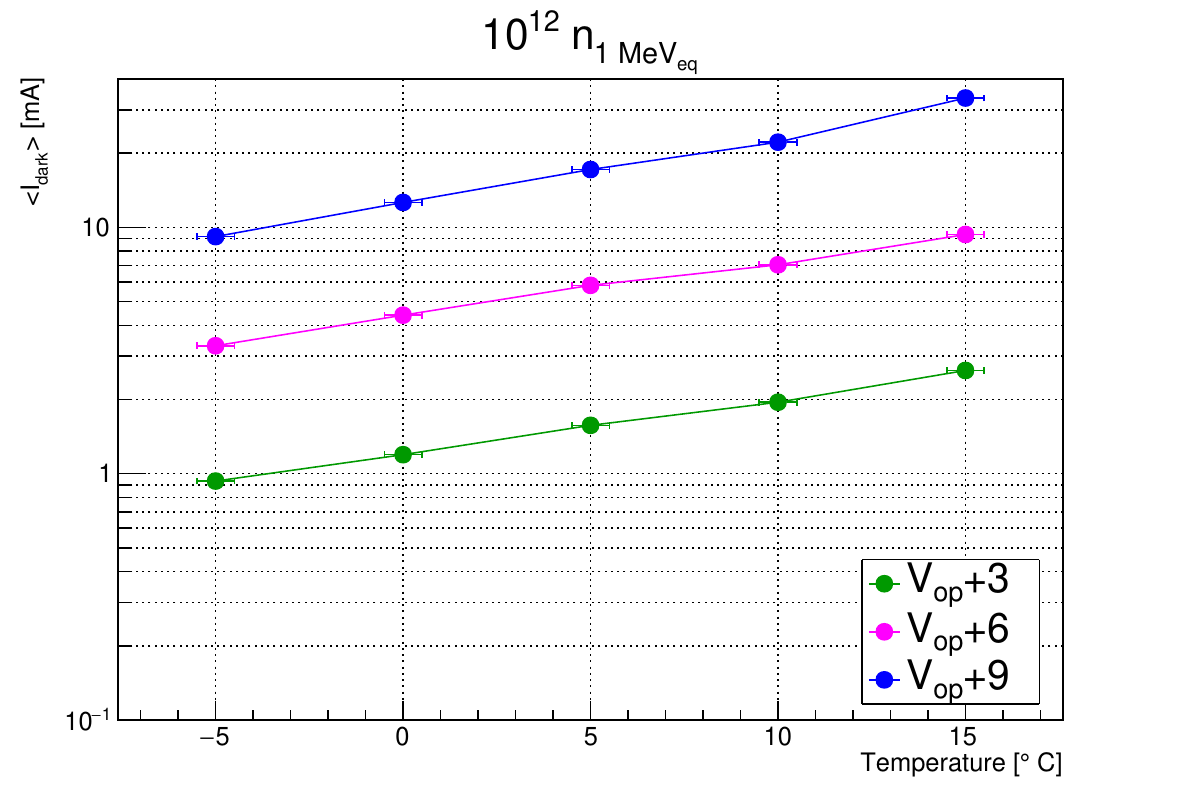}
    \caption{Left: Mean value of the leakage current as a function of the SiPMs temperature for SiPM irradiated at different neutron fluences (red curve: $10^{12}$ n$_{1MeV_{eq}}$,
    blue curve: $5 \times 10^{11}$ n$_{1MeV_{eq}}$, green curve: $10^{11}$ n$_{1MeV_{eq}}$, black curve: $5 \times 10^{11}$ n$_{1MeV_{eq}}$). Right: $I_{dark}$ dependence on SiPM temperature for different overvoltages.}
    \label{fig:irradiation}

\end{figure}
In \figurename~\ref{fig:irradiation}, some of the results obtained with the SiPMs irradiated at FNG are reported. Among all the sample, we conservatively selected only samples with higher leakage current.
%the data acquired in the two different facilities present a $\mathcal{O}(1.5)$ mA difference in terms of increase leakage current at operational voltage.
In \figurename~\ref{fig:irradiation} (Left) the leakage currents  measured at $V_{op}$ at different temperatures and fluences are reported: the dependence of $I_{dark}$ to the sensor temperature appears to be linear and a factor of $\sim 2$ decrease in leakage current is observed when reducing the temperature by $+10 ^{\circ}$ C. Looking at \figurename~\ref{fig:irradiation} (Right), that represents the leakage current dependence on temperature at fluence compatible to the lifetime of the experiment, for different bias voltages ($V_{br}+3$,$V_{br}+6$, $V_{br}+9$), it is clear that the calorimeter will need to operate the SiPMs down to $-10 ^{\circ}$ C, in order to meet the 2 mA limit on each channel, assuming only a small decrease in $V_{bias}$.

\subsection{Calorimeter assembly at Fermilab}
%%%%%%%%%%%%%
According to the current schedule, it is planned that the calorimeter will be installed in the experimental hall by summer 2024 and all the service lines will be completed by the autumn of that same year. Soon after, a commissioning data-taking with cosmic rays will start, while waiting for the tracker to be installed.  To respect this schedule, the disks have to be assembled by the spring of 2023. 

\begin{figure}[h!]
    \centering
    \includegraphics[width = 0.46 \textwidth]{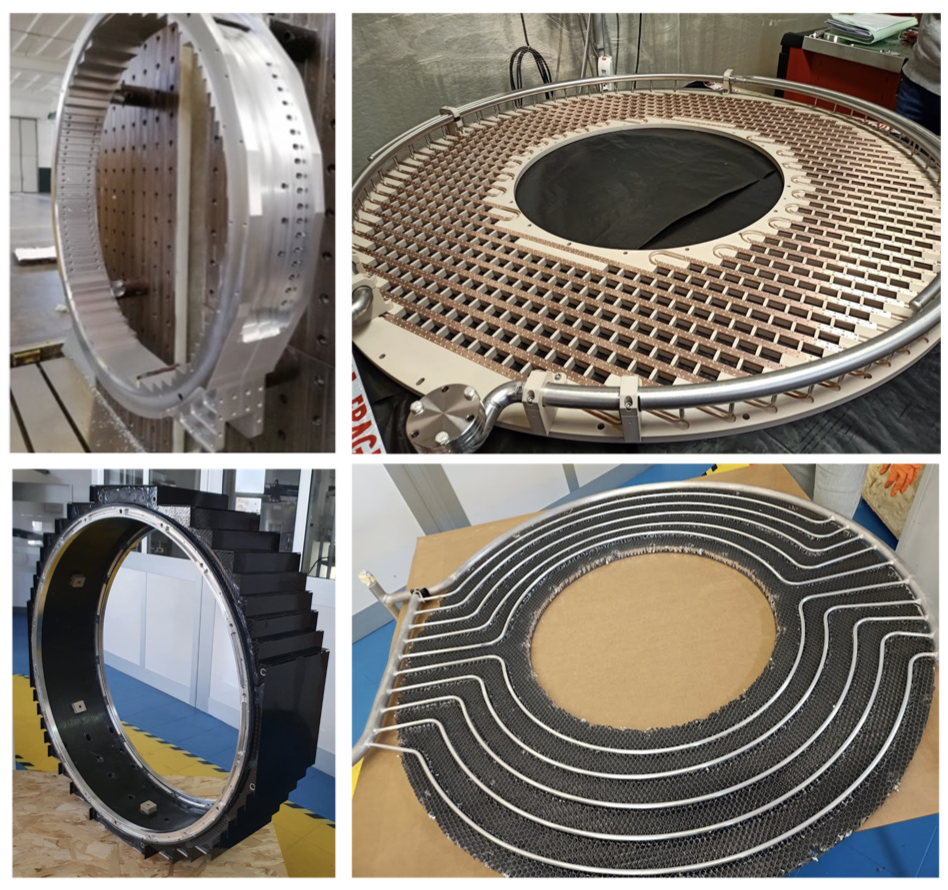}
    \caption{ Calorimeter mechanical parts: (Top-left) Outer Ring, (Top-right) FEE plate, (Bottom-left) Inner Ring, (Bottom-right) Front plate with source tubing embedded in the aluminum honeycomb.}
    \label{fig:meccanica}
\end{figure}

Within the last two years, all large mechanical parts (see \figurename~\ref{fig:meccanica}) have been produced. Before shipping them to Fermilab, a dry-fit was carried out in a clean room at LNF to check that all pieces were well fitting as shown in \figurename~\ref{fig:meccanica}. A careful vacuum leak test was performed both on the crate manifolds and on the elbow joints between crates and manifolds: a maximum leak rate of $10^{-10}$ atm $\times$ cc/s was achieved, satisfying the experiment requirements.

\begin{figure}[h!]
    \centering
    \includegraphics[width = 0.48 \textwidth]{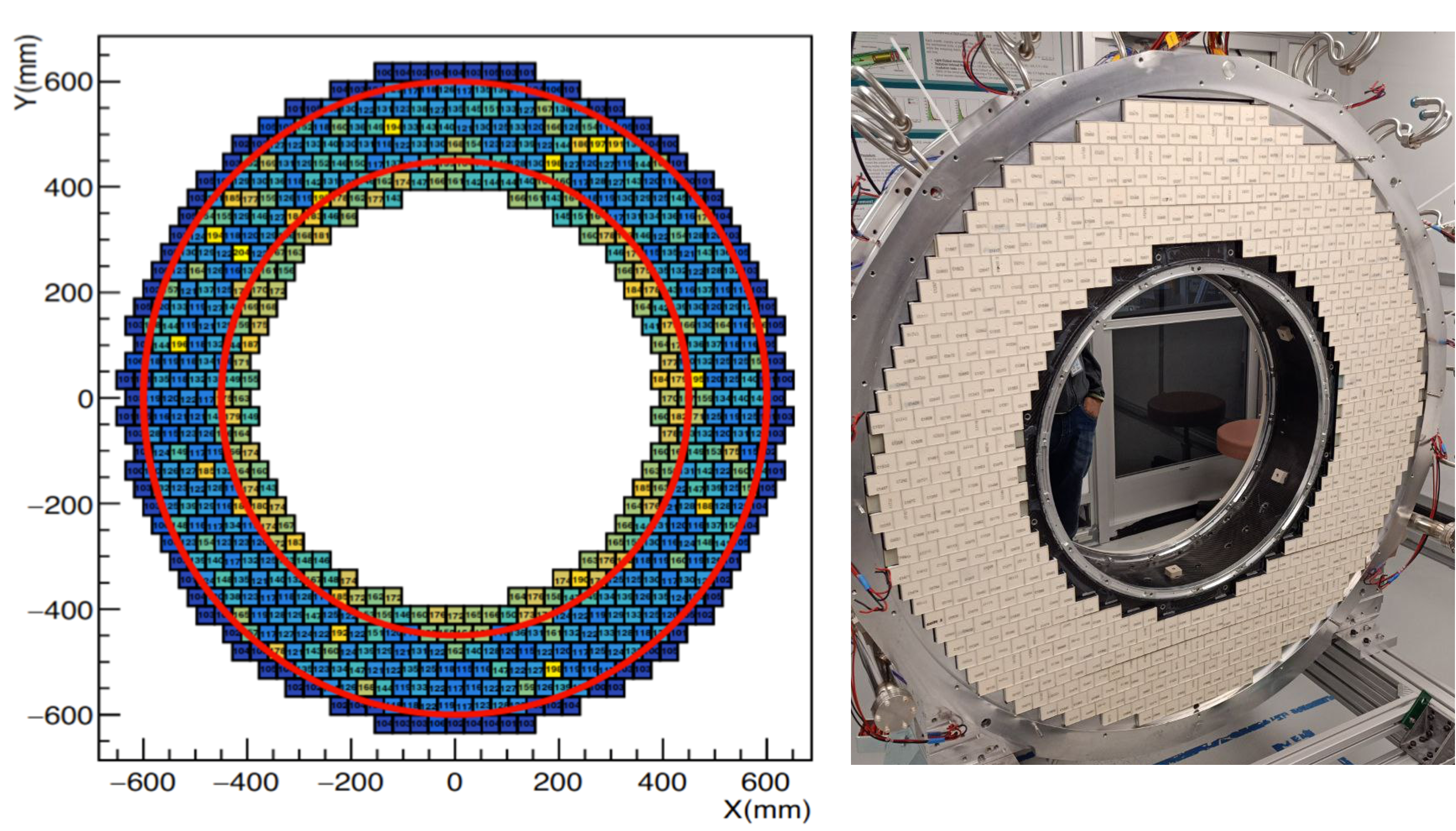}
    \caption{Left:Distribution of the LY in photoelectrons/MeV and down-stream disk filled with 674 crystals}
    \label{fig:disk1}
\end{figure}

At the moment, the down-stream disk (Disk1) is under assembly at the Silicon Detector facility (SiDet) at Fermilab in a clean room (ISO 7), while the up-stream disk (Disk0) has all the mechanical parts assembled at LNF and is still waiting to be shipped to Fermilab.
The Disk1 assembly at SiDet started in June 2022 with the installation and alignment of the Outer Ring over its stand, and progressed with the Inner Ring alignment.
674 crystals, among the over 1450 wrapped in Tyvek\textregistered{}  and fully characterized with a source, were selected to be stacked in the down-stream disk. Before stacking, a set of two day long outgassing runs was performed in  a dedicated vacuum vessel to reduce the residual single crystal outgassing level below $10^{-4}$ Torr l/s.
The positioning of the crystals in the disk was carefully optimised, taking into account optical, mechanical and radiation hardness properties. Three areas have been identified as a function of the expected level of Total Ionizing Dose (TID), occupancy and resolution, considering that TID and occupancy decrease radially moving toward the outermost part of the disk. The chosen crystal distribution is shown in Figure~\ref{fig:disk1}.

A 50 $\mu$m thick Tedlar\textregistered{} layer was placed between each stacked row; to keep the crystal matrix solidly connected, each row was  then compressed using screws pushing on plastic shims at both row ends. A high precision bubble level allowed to check the accuracy of stacking procedure, that was later reconstructed offline after an overall survey with a laser tracker.

Starting from fall 2022, the front panel with embedded source was installed and coupled to the Carbon Fiber Inner Ring and the outer Al disk. Subsequently the ten crates supporting the mezzanine and Dirac board, as well as the connection to the cooling circuits, were installed and the possible losses were carefully tested with a Helium sniffer, proven to be lower than $10^{-11}$ Torr $\times$l/s.

\begin{figure}[h!]
    \centering
    \includegraphics[width = 0.4 \textwidth]{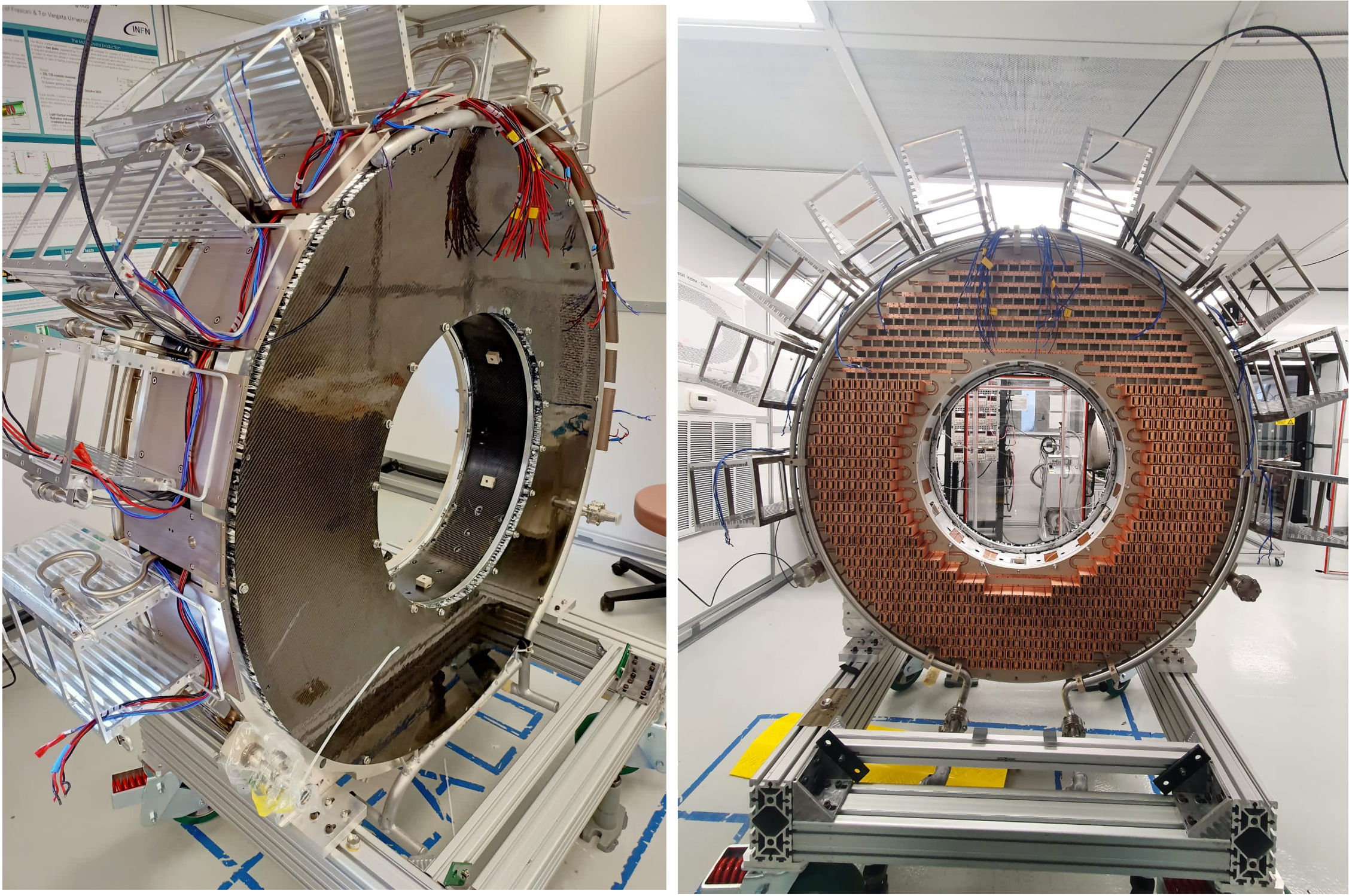}
    \caption{Front and rear face of the down stream disk of the Mu2e calorimeter}
    \label{fig:finalStatus}
\end{figure}

The first 500 ROUs were received at Fermilab in September 2022.
An outgassing campaign of these ROUs  started in October 2022  to reduce their outgassing level below $10^{-4}$ Torr l/s before inserting them in the FEE disk. In \figurename~\ref{fig:finalStatus} the current status of the assembly is shown.

\section{Conclusion}
In this paper the construction status of the Mu2e electromagnetic calorimeter is summarized. The detector is composed of two annular disks, each filled with 674 pure CsI crystals. Each crystal is coupled to two custom UV-extended SiPMs, to provide redundancy.\\
While the production of crystals, SiPMs and FEE boards has been completed, the digital electronics production is still underway.  
The assembly of the ROU,  each one composed by two SiPMs and two FEE units, is more than 70\% complete and a full characterization of the gain, and of the gain-dependence on bias voltage, is being successfully carried out. This calibration will allow to adjust the calorimeter response
once in operation and after the detector will be exposed to a large neutron fluence. Knowing this behavior ``apriori'' will be pivotal when the neutron fluence will cause an increase of the SiPMs dark current.
To keep the leakage current of each channel under the 2 mA limit set by the read-out board, it will be possible either to decrease the operating temperature of the SiPMs or reduce the bias voltage.\\
At the moment of writing all mechanical parts of the calorimeter have been delivered and the downstream disk is under assembly at Fermilab.
The calorimeter support structure has been completed and all crystals stacked. The mounting of the Read-Out Units is started with more than 2/3 being inserted in their own position in the readout plate. It is expected to conclude this part of the assembly by the end of 2022 and then repeat the same steps for the other disk.\\
Finally, as soon as the digital electronics will be delivered, we plan to complete an integrated test of the calorimeter readout starting from summer 2023 before getting ready to move the calorimeters in the experimental hall.

\section*{Acknowledgements}
We are grateful for the vital contributions of the Fermilab staff and the technical staff of the participating institutions. This work was supported by the US Department of Energy; the Istituto Nazionale di Fisica Nucleare, Italy; the Science and Technology Facilities Council, UK; the Ministry of Education and Science, Russian Federation; the National Science Foundation, USA; the National Science Foundation, China; the Helmholtz Association, Germany; and the EU Horizon 2020 Research and Innovation Program under the Marie Sklodowska-Curie Grant Agreement Nos.  734303, 822185, 858199, 101003460, and 101006726. This document was prepared by members of the Mu2e Collaboration using the resources of the Fermi National Accelerator Laboratory (Fermilab), a U.S. Department of Energy, Office of Science, HEP User Facility. Fermilab is managed by Fermi Research Alliance, LLC (FRA), acting under Contract No. DE-AC02-07CH11359.

\end{document}